\documentclass[aps,prb,showpacs,twocolumn]{revtex4}

\newcommand{\DEs}{\Delta E_{\scriptscriptstyle\rm S}}
\newcommand{\muv}{\boldsymbol \mu}
\newcommand{\deltav}{\boldsymbol \delta}
\newcommand{\Mv}{\mathbf M} 
\newcommand{\Kv}{\mathbf K}
\newcommand{\kv}{\mathbf k} 
\newcommand{\Rv}{\mathbf R}
\newcommand{\Ux}{U_{\rm x}} 
\newcommand{\Ut}{U_{\rm t}}
\newcommand{\kB}{k_{\scriptscriptstyle\rm B}}

\usepackage{graphics} 
\usepackage{amsmath}
\usepackage{epsfig}

\begin{document}

\title{Epitaxial growth of binary alloy nanostructures}

\author{S. Heinrichs,$^1$ W. Dieterich,$^1$ and P. Maass$^2$}

\affiliation{$^1$Fachbereich Physik, Universit\"at Konstanz D--78457
  Konstanz, Germany} \affiliation{$^2$Technische Universit\"{a}t
  Ilmenau - Institut f\"ur Physik, D--98684 Ilmenau, Germany}

\date{July 11, 2006}

\begin{abstract}
  Stochastic growth of binary alloys on a weakly interacting substrate
  is studied by kinetic Monte Carlo simulation. The underlying lattice
  model relates to fcc alloys, and the kinetics are based on
  deposition, atomic migration with bond-breaking processes and
  exchange processes mediated by nearest neighbor hopping steps. We
  investigate the interrelation between surface processes and the
  emerging nonequilibrium structure at and below the growing surface
  under conditions where atoms in the bulk can be regarded as
  immobile. The parameters of the model are adapted to CoPt$_3$
  alloys. Growing nanoclusters exhibit an anisotropic short range
  order, primarily caused by Pt segregation at the surface. The
  overall structural anisotropy depends on both Pt surface segregation
  and cluster shape, and can explain the perpendicular magnetic
  anisotropy (PMA) recently measured in CoPt$_3$ nanoclusters on a van
  der Waals substrate.  The onset of L1$_2$ ordering in the cluster is
  induced by surface processes. The same kinetic model is applied also
  to continuous thin films, which in addition can exhibit a small bulk
  contribution to PMA.
\end{abstract}


\maketitle

\section{Introduction}
\label{sec:introduction}

Molecular beam epitaxy (MBE) has become an important tool to prepare
ultrathin films and nanostructures in nonequilibrium, frozen-in states
that show novel properties distinctly different from the equilibrium
bulk phases. Theoretically, an important question is how to relate the
structural characteristics of the growing material to the incoming
flux, substrate temperature, adatom diffusion coefficients and adatom
interactions. Successive stages of growth, nucleation and island
formation in the submonolayer regime,\cite{Venables+84} second layer
nucleation,\cite{Rottler+99,Heinrichs+00} growth of three-dimensional
clusters and emerging film morphologies, are fairly well understood
for the one-component case, i.e.\ for deposition of only one species of
atoms.\cite{Brune98,Michely+04} By contrast, nonequilibrium alloy
nanostructures produced by codeposition of two or more atomic species
can display a variety of additional new phenomena that are only
partly understood up to now. From the viewpoint of theoretical
modelling, this is true in particular for metallic nanoalloys, which
recently became of great interest from the experimental and
technological viewpoint.\cite{Albrecht+02}

In many MBE experiments the substrate temperature is chosen
sufficiently low so that atomic configurations in the interior of a
growing cluster or film can be considered as frozen. Structural
relaxation then takes place only among lower-coordinated atoms within
the growth zone. Specific to binary systems are the following
questions:
\begin{itemize}
\item[{i)}] Suppose that one atomic species tends to segregate at the
  surface. What is the emerging bulk structure upon further deposition
  burying a segregated surface layer?
  
\item[{ii)}] How large is a possible anisotropy in the frozen
  short-range order, i.e.\ a difference in the alloy structure between
  the lateral and the perpendicular (growth) direction?
  
\item[{iii)}] To what extent can diffusion processes limited to the
  near-surface region give rise to long-range order in the bulk of an
  ordering alloy at temperatures below the equilibrium order-disorder
  transition point?
  
\item[{iv)}] In the case of magnetic alloys, what are the magnetic
  properties associated with the non-equilibrium short or long-range
  order? This question is central in the ongoing search for materials
  that display a stable perpendicular magnetic anisotropy (PMA), which
  is useful for the development of high-density magnetic storage
  media.  Under this viewpoint fcc-alloy systems of
  Co-Pt,\cite{Albrecht+01,Shapiro+99} Fe-Pt\cite{Zeng+02} or
  Fe-Co\cite{Andersson+06} draw great attention in current
  investigations.
\end{itemize}

Here we report on answers related to these problems, based on kinetic
Monte Carlo (KMC) simulations of a statistical model for binary fcc
alloys of composition AB$_3$, with emphasis on CoPt$_3$ alloys.

MBE-grown nanoclusters of CoPt$_3$ on a van der Waals WSe$_2$
substrate were found recently to exhibit PMA at room temperature and
below the onset of L1$_2$-ordering.\cite{Albrecht+01} The origin of
PMA is thought to lie in a structural anisotropy induced by the growth
process, strong enough to overcome the dipolar form anisotropy
favoring in-plane magnetization. The model we use is based on
effective chemical interaction and kinetic parameters extracted,
respectively, from known equilibrium properties of
CoPt$_3$\cite{Sanchez+89,Gauthier+92} and from separate diffusion
measurements.\cite{Bott+96} The dependence of the magnetic anisotropy
on the local structure is represented by a sum over bond
contributions, with nearest neighbor Co-Pt bonds as the dominant part.
The associated bond anisotropy parameter is deduced from magnetic
measurements on Co-Pt multilayer systems.\cite{Johnson+96} Within this
model it was shown that Pt surface segregation and cluster shape
effects can explain the occurence of PMA in some temperature window,
followed by the evolution of L1$_2$-order at higher
temperatures.\cite{Heinrichs+06} In this paper we present further
details on cluster structures and extend our previous investigations
to homogeneous films.

In section~\ref{sec:model} we introduce our kinetic model, based on
random deposition and subsequent hopping moves of A and B atoms to
neighboring sites of an fcc lattice. Hopping rates are derived from
the energetics in the specific environment of each atom. To our
knowledge, most prior KMC simulations of three-dimensional growth of
binary systems were in the framework of solid on solid (SOS)
models,\cite{Barabasi+95} which are not well suitable to describe
cluster shapes and facetting in a realistic way. In very recent work
the growth of CoPt$_3$ films was simulated, using a more realistic
tight binding, second moment approximation to the interatomic
interaction potential.\cite{Maranville+06} Segregation of Co-atoms to
step edges on the growing surface was predicted, which leads to
in-plane Co-clustering.\cite{Cross+01} By this the authors were able
to explain PMA in disordered films, known from previous
measurements\cite{Shapiro+99} to occur at elevated temperatures ($T
\gtrsim 500$K).

Our model, based on effective nearest neighbor couplings, allows for
fully three-dimensional kinetics, a feature which is important in
simulations of cluster growth and L1$_2$ ordering. It yields cluster
facets in agreement with those observed in experiments and
consistently accounts for mass transport along the cluster side and
top facets. Both height and lateral growth largely relies on
interlayer diffusion that starts from adatoms on the substrate,
reaching higher layers along side facets. The present model
automatically entails step edge barriers, dissociation from the
surface and the possibility of vacancy diffusion in the bulk. It is
less comprehensive than recent simulations of island formation in
compound semiconductors, which are based on extensive electronic
density functional calculations for energy barriers of the processes
involved. These calculations, however, were so far limited to the
submonolayer regime.\cite{Kratzer+02} Section~\ref{sec:magnetic} also
describes a phenomenological relationship between the local
contributions to the magnetic anisotropy and atomic short range order
within a bond picture.

The anisotropic short range order and associated magnetic properties
of clusters up to 4000 atoms are presented in
section~\ref{sec:clusters}, and are discussed in detail in terms of
surface segregation and cluster shape. The same algorithm is applied
in section~\ref{sec:films} to thin films. In addition to surface
-induced PMA a bulk contribution is identified which, however, is too
small to account for the measured PMA in thick films. Further
conclusions are drawn in section~\ref{sec:conclusions}.

\section{Simulation model}
\label{sec:model}
\subsection{Interactions and Kinetics}
\label{sec:interactions}
Our model starts from a reference fcc lattice with nearest neighbor
distance $a$ in a cubic simulation box $L\times L\times L_z$ where
sites can be occupied by an atom of type A (Co), B (Pt), or by a
vacancy V. In contrast to models with a small concentration of
vacancies and fixed shape of the sample, most of the simulation box
considered here consists of vacancies representing the free space in
the vicinity of the cluster. This allows for an unconstrained
evolution of the cluster morphology. Effective interactions between A
and B atoms are restricted to nearest neighbor pairs.  This simplified
description already captures essential statistical properties of
ordering fcc alloys, including the phase diagram which displays L1$_2$
and L1$_0$ ordered structures in the vicinity of the AB$_3$ and
AB-composition, respectively.\cite{Sanchez+89,Kessler+01_03}

Bond energies between the different atomic species are denoted by
$V_{\rm AA}, V_{\rm AB}$ and $V_{\rm BB}$. The linear combinations $I
= \frac{1}{4}(V_{\rm AA} + V_{\rm BB} - 2 V_{\rm AB})$, $h = V_{\rm
  BB} - V_{\rm AA}$ which acts as a surface field, and $V_0 =
\left(V_{\rm AB} + V_{\rm BB}\right)/2$ control the bulk
order-disorder transition temperature, the degree of surface
segregation of B-atoms, and the average bond energy in the L1$_2$
ordered state. Parameter values are adjusted to reproduce equilibrium
properties of CoPt$_3$: (i) The transition from the disordered to the
L1$_2$-structure occurs at $T_0 \simeq 958\,$K,\cite{Berg+72} which is
related to $I$ by $k_B T_0 \simeq 1.83I$.\cite{Binder80} Setting $I =
1$, our energy unit becomes $k_B T_0/1.83 = 45\,$meV, corresponding to
523$\,$K. (ii) The observed Pt surface segregation of nearly
100\%,\cite{Gauthier+92} caused by the larger size of Pt relative to
Co, is compatible with $h \gtrsim 4$; here we choose $h = 4$ in most
of our calculations. (iii) At temperatures of interest, the mobility
of atoms is effectively restricted to the film or cluster surface so
that their typical coordination is between 3 (for an adatom on top of
a terrace) and 7 (for an atom attached to a step edge). Parameters for
a variety of corresponding processes were calculated for Pt within
electronic density functional theory (DFT).\cite{Feibelman99}
Different processes were classified according to the number of broken
bonds, and the average bond energy yielded $V_0 \simeq - 5$, which we
use here. Clearly, these bond energies are to be understood as
effective energies since real interactions can extend to several
neighbors.\cite{Kentzinger+00}

Experiments for nanoclusters of CoPt$_3$ were performed on a WSe$_2$
(0001) substrate surface,\cite{Albrecht+01} which is of the van der
Waals type. The interaction with the substrate is modelled by a
weak attractive potential, represented by an additional energy $V_s$
for atoms in the first layer. In the cluster simulations we choose
$V_s=-5$. Compared to an fcc Pt (111) surface with three bonds of
typical strength $V_0$, this amounts to about 1/3 of the energy of a
single Pt-Pt bond.

The total flux $F$ of incoming atoms is taken as $F = 3.5$
monolayers/s.  Additional parameters needed to describe the kinetics
of the system are the transition state energy $\Ut$ and the attempt
frequency $\nu$ for atomic moves. These values are known from
diffusion experiments\cite{Bott+96} of Pt on Pt (111) surfaces as $\nu
= 8.3 \cdot 10^{11}$ s$^{-1}$ and $\Ut \simeq 5$. The resulting
diffusion coefficient is $D/a^2 = (\nu/4) e^{- \Ut/k_B T}$ for moves
without change in energy.

The rate for an elementary hopping process with a final energy $E_f$
and initial energy $E_i$ is chosen to be
\begin{equation}\label{eq:hopping_rate}
w_{if} = \nu e^{- \Ut/k_B T} \min (1, e^{-(E_f - E_i)/k_B T})
\end{equation}
which fulfills the condition of detailed balance. As long as the
general kinetics are concerned, characterized by the tendency of the
system to approach thermal equilibrium, the specific form
(\ref{eq:hopping_rate}) of the hopping rates should not be of crucial
importance.

In many cases,\cite{Gambardella+00,Santis+02} including Co deposited
on Pt(111), adatoms on the top of a terrace or at step edges can
exchange positions with an atom underneath in one concerted move. In
comparison to single atom moves considered so far, a larger binding
energy has to be overcome in order to effect such a simultaneous move
of two particles. However, at the same time the transition state
energy can be reduced because of the higher coordination in the
transition state configuration. Direct exchange processes are found
for Co deposited on Pt(111) over a wide temperature range
250--520$\,$K.\cite{Gambardella+00,Santis+02} They are especially
frequent for low-coordinated atoms on top of terraces (with
coordination 3) or at step edges (with coordinations 4 or 5). We
therefore include direct exchange processes between unlike atoms in
the simulation. The corresponding rate involves an exchange barrier
$\Ux$ that adds to the migration barrier $\Ut$, while initial and
final energies are taken into account as before. These exchange
processes, however, are allowed only if one atom has a coordination in
the range 3 -- 5 and the other one in the range 8 -- 10. Let us remark
that the activation energy for exchange diffusion of Pt on Pt (100) is
just 470$\,$mV,\cite{Kellog94} which is almost the same as the total
barrier $\Ut + \Ux$ when $\Ux = 5$, as used in most of our
computations described below.

The time evolution in our simulation is determined by a rejectionless
continuous time MC-algorithm that generates realizations of the master
equation describing the growth kinetics. Each simulation step consists
of: (i) incrementing the time by an interval $\Delta t$ drawn from an
exponential distribution with mean $\langle \Delta t \rangle = (\sum_f
w_{if})^{-1} $, (ii) executing a process $i \rightarrow f$ with
probability proportional to its rate $w_{if}$, (iii) updating of rates
of processes affected by the moving atom.

\begin{figure}
\begin{center}
  \includegraphics[width=7.3cm]{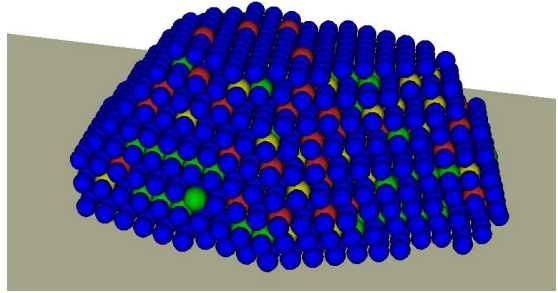}
\end{center}
\caption{A typical cluster configuration with 2000 atoms for $h=4$,
  $T=1.2$ and $\Ux=0$. Pt atoms are marked in blue and Co atoms are
  marked in different colors depending on the number of nearest
  neighbor Pt atoms: if more (less) Pt atoms are found out of plane
  than in plane the atom is marked in green (yellow); otherwise it is
  marked in red.  Facets with sixfold \{111\} and fourfold \{100\}
  symmetry can be clearly distinguished.}
\label{fig:cluster-shape}
\end{figure}
The simulation box consists of 15 layers with 1440 lattice sites each.
To generate an isolated cluster attached to the weakly binding
substrate, we start from a seed with 5 atoms on the surface.  For the
deposition rate $F=3.5$~ML/s used in most simulations and temperatures
$T>0.5$, the distance between clusters, which would be observed by
self-organized nucleation, exceeds the box size. This could be
effectively accounted for by increasing the deposition rate at the
boundary of the box to match the extra flux of atoms deposited within
the typical capture zone of an island. However, in order to reduce the
number of parameters, we deliberately kept the deposition rate
constant for all temperatures.

An example of a typical cluster with $N=2000$ atoms obtained in the
KMC simulations is shown in Fig.~\ref{fig:cluster-shape}.  The top
facet has (111) orientation as expected for the (111) substrate, and
side facets are of (111) and (100) type. The surface shows strong Pt
segregation. Similar to the experiments,\cite{Albrecht+02} the aspect
ratio of lateral to height dimension of the cluster is about 3:1. At
the temperature $T = 1.2$, the ratio $D/F a^4 \simeq 2 \cdot 10^{10}$,
and the realization of the growth process implied $3.3 \cdot 10^{10}$
elementary processes. A simple MC algorithm which generates equivalent
dynamics, but is not rejection free, can be estimated to require about
$N^2 D/(a^4 L^2 F)\simeq 6\cdot 10^{13}$ trials for elementary moves,
i.e.\ about 2000 times more than the algorithm used here. This ratio
increases further at lower temperatures, since bond breaking processes
will acquire lower rates and consequently contribute to larger time
increments $\Delta t$. Subsequent equilibration of the clusters under
zero flux during a multiple of the deposition time did not induce
significant changes in the cluster configurations and will therefore
be ignored in the following.

To generate homogeneous films (where the width of the interfacial
growth zone is much smaller than the average film thickness),
discussed in section~\ref{sec:films}, we choose a stronger surface
potential $V_s = -15$ which guarantees complete wetting in the first
layer in the temperature range considered.

\begin{table}
\caption{Parameters used in the simulation. All energies are given in
  units of $k_B T_0/1.83 = 45\,$meV.}
\begin{tabular}{@{\hspace{0.3cm}}c@{\hspace{0.3cm}}|@{\hspace{0.3cm}}c@{\hspace{0.3cm}}|@{\hspace{0.3cm}}c@{\hspace{0.3cm}}|@{\hspace{0.3cm}}c@{\hspace{0.3cm}}||@{\hspace{0.3cm}}c@{\hspace{0.3cm}}|@{\hspace{0.3cm}}c@{\hspace{0.3cm}}|@{\hspace{0.3cm}}c@{\hspace{0.3cm}}}
\hline
\hline
$I$ & $h$ & $V_0$ & $V_s$ & $\Ut$ & $\Ux$ & $\nu$                     \\\hline
1   & 4  & -5    & -5    & 5     & 5     & $8.3\cdot 10^{11} s^{-1}$  \\
\hline
\hline
\end{tabular}
\end{table}

\subsection{Structure-induced magnetic properties}
\label{sec:magnetic}

In CoPt$_3$-alloys magnetic moments are mostly due to the Co atoms,
with $\mu^{\rm Co} \simeq 1.7 \mu_B$,\cite{Menzinger+66,Shapiro+99}
while Pt atoms carry a comparatively small induced moment $\mu^{\rm
  Pt}$ of about $0.3\mu_B$, which has been shown to depend on the
actual atomic environment.\cite{Sanchez+89} Hybridization of
d-electrons between neighboring Co and Pt atoms and a strong
spin-orbit interaction near the Pt sites lead to a magnetic anisotropy
that tends to align the Co moment along the Co-Pt bond direction. In
the following we adopt a bond picture as in earlier work on bulk
systems,\cite{Neel54,Victoria+93} where the structural part of the
magnetic anisotropy, $H_A$, is expressed as a sum over bonds $\langle
i, \delta\rangle$ that connect a Co atom with moment $\muv_i ^{\rm
  Co}$ at site $\Rv_i$ with a species $\alpha$ ($\alpha$ = Co,Pt,V) at
a site $\Rv_i + \deltav$,
\begin{equation}\label{Ha}
  H_A = - \sum _{\langle i, \delta\rangle} \sum_\alpha A^{\rm Co
  \, \alpha} (\muv_i ^{\rm Co}\cdot \deltav)^2/(|\muv_i
  ^{\rm Co} | |\deltav|)^2
\end{equation}
By $\deltav$ we have denoted the 12 possible nearest neighbor bond
vectors in the fcc lattice. Note that this anisotropy term naturally
entails both surface and bulk contributions by considering vacancies
as a possible neighbor species.

The parameters $A^{\rm Co \, \alpha}$ are the anisotropy energies
associated with a Co-$\alpha$ bond. These are deduced from experiment
and the term $\alpha=$ Pt gives the dominant contribution. Note that
for saturated magnetization a nearest neighbor contribution to the
anisotropic part of dipole-dipole interactions has exactly the same
form, so that this contribution is included already in the
coefficients $A^{\rm Co\,\alpha}$. The isotropic part of the exchange
interactions only leads to a small renormalization of chemical
interactions.

Equation~(\ref{Ha}) provides a relationship between a given atomic
structure, to be obtained from simulations, and the magnetic
anisotropy energy. Magnetic anisotropies for disordered alloys,
including Co-Pt alloys, have been obtained by microscopic calculations
based on the KKR-CPA method.\cite{Staunton+98} Although less accurate
in principle, the local expression (\ref{Ha}) here allows us to relate
effects in the atomic short range order to the anisotropy energy in a
direct way.  Note that 2nd order contributions of the anisotropy in
the direction cosines as in eq.~(\ref{Ha}) do not yield a
magneto-crystalline anisotropy in an fcc lattice occupied by a single
species. Additional bulk contributions are of 4th order and generally
much weaker than the magnetic anisotropy in lower symmetry
configurations to be considered here.

Several observations support the bond picture underlying
eq.~(\ref{Ha}) as a reasonable approximation. Magnetic torque and
magneto-optical Kerr-effect measurements on Co-Pt multilayers show
that the anisotropy energy in these systems is dominated by an
interfacial contribution connected to Co-Pt bonds.  Detailed
measurements of multilayers yield $K^{\rm CoPt}=0.97$~mJ/m$^2$ for
(111) orientation and $K^{\rm CoPt}=0.59$~mJ/m$^2$ for (100)
orientation.\cite{Weller+93,Johnson+96} Considering the different
angles of bonds to the surface and the different packings,
eq.~(\ref{Ha}) indeed can reproduce this difference with one
consistent value for $A^{\rm CoPt} \approx 250\,\mu$eV per Co-Pt
bond.\cite{comm-A_par} Hence, on a semiquantitative level, anisotropy
energies of Co-moments in different chemical environments are
consistent with a superposition of bond contributions and a bond
energy $A^{\rm CoPt}$ as given above.

Similarly, from measurements on Co-vacuum
interfaces\cite{Kohlhepp+95,Beauvillain+94} and theoretical
investigations of a freely standing Co-monolayer\cite{Wu+91} we
estimate $A^{\rm CoV} \simeq -67\,\mu$eV. The remaining parameter is
$A^{\rm CoCo}$ for which we retain only the nearest neighbor dipolar
contribution as discussed above. Using $\mu^{\rm Co} = 1.7 \mu_{\rm
  B}$ and $|\deltav| = 2.72$~\AA\ in CoPt$_3$, this yields $A^{\rm
  CoCo}=23\,\mu$eV.

It is now straightforward to express the part of the anisotropy energy
caused by local chemical order,
\begin{equation}\label{anisotropy}
E_s = H_A\{\muv {\rm{\,  in\,  plane} } \} - H_A \{\muv
{\rm {\, out\,  of\,  plane}}\}
\end{equation}
in terms of the numbers of Co-$\alpha$ bonds in plane,
$n_\parallel^{{\rm Co} \, \alpha}$, and out of plane, $n_\perp^{{\rm
    Co} \, \alpha}$. Using a site occupation number representation and
the condition that occupation numbers for the three species $\alpha$
add up to unity at each site, (\ref{anisotropy}) can be brought into
the form
\begin{equation}\label{Es}
E_s = \sum_{\alpha = {\rm Co, Pt}} E_s^{\rm Co\,\alpha}=\frac{N}{2}\sum_{\alpha = {\rm Co, Pt}}(A^{{\rm Co} \,
\alpha} - A^{\rm Co V} ) P^{{\rm Co }\, \alpha}
\end{equation}
with structural anisotropy parameters
\begin{equation}\label{parameter}
  P^{{\rm Co}\,\alpha} = \frac{1}{N}(n_\perp^{{\rm Co}\,\alpha} -
  n_\parallel^{{\rm Co}\,\alpha}).
\end{equation}
These parameters enter as the primary structural characteristics
related to the magnetic anisotropy.

In addition one has to take into account the magnetic form anisotropy
due to dipolar interactions. Dipolar sums for a given cluster with
saturated magnetic moment $\Mv_s$ are carried out using moments for
Co and Pt as given before.  By $E_{\rm dip}$ we denote the difference
in dipolar energies for $\Mv_s $ parallel and perpendicular to the
substrate, respectively. The anisotropy within the substrate plane
turns out to be negligibly small because of the approximate 3-fold
symmetry of cluster shapes. Nearest neighbor dipole-dipole
interactions were already incorporated in the quantity $E_s$, as
described above. The total magnetic anisotropy energy of a cluster is
then given by
\begin{equation}\label{Etot}
  E_{\rm tot} = E_s + E_{\rm dip}
\end{equation}
Dipolar interactions generally favor in plane magnetization of thin
films and clusters, $E_{\rm dip} < 0$. For thin homogeneous films,
$E_{\rm dip} = - (\mu_0/2) M_s ^2$. 

As mentioned above and supported by the above estimates, the term
$\alpha = $Pt dominates expression (\ref{Es}). PMA hence requires
$P^{\rm CoPt}$ to be sufficiently large that $E_s$ can overcome the
negative dipolar energy.

\section{Clusters}
\label{sec:clusters}
\begin{figure}
\begin{center}
  \includegraphics[width=9cm]{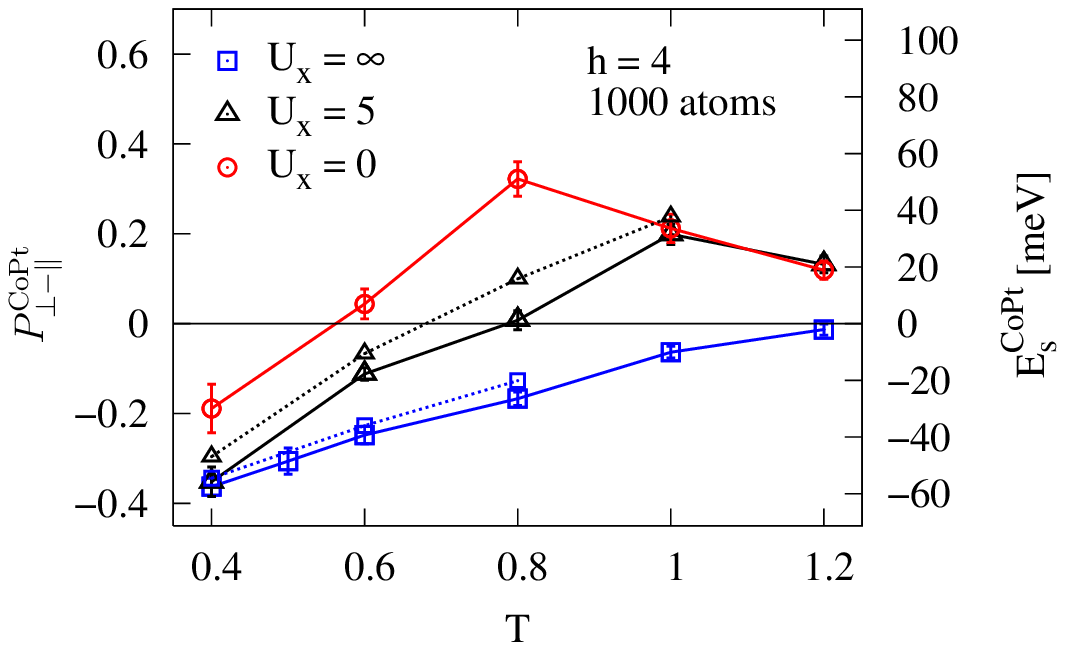}\\
  \includegraphics[width=9cm]{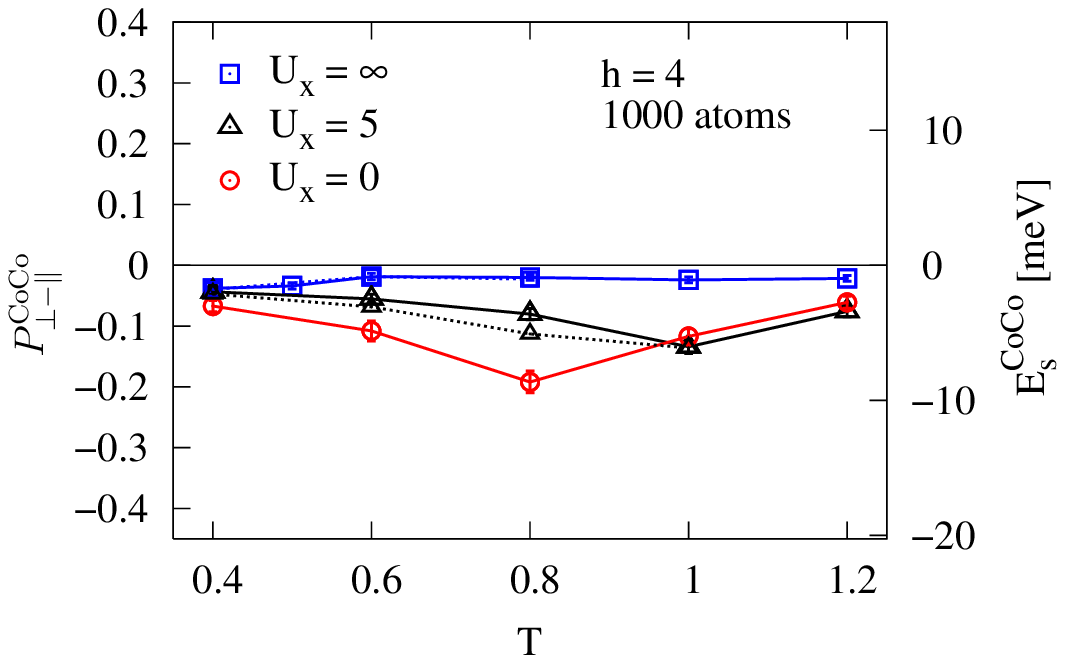}
\end{center}
\caption{Structural anisotropy parameters $P_{\perp-\|}^{\rm
    Co\!-\!Pt}$ (upper panel) and $P_{\perp-\|}^{\rm Co\!-\!Co}$
  (lower panel) and corresponding structural anisotropy energies. The
  dashed lines correspond to data obtained with a deposition rate of
  $F=0.35$~ML/s, which is 10 times smaller than the rate used for the
  full lines.}
\label{fig:struct_anisotropy-T}
\end{figure}

The two anisotropy parameters $P^{{\rm CoPt}}$ and $P^{\rm CoCo}$
defined in (\ref{parameter}) are shown in
Fig.~\ref{fig:struct_anisotropy-T} as a function of temperature for
clusters consisting of 1000 atoms. The solid lines connect data points
for different values of the extra barrier for exchange processes
$\Ux$, and for a flux $F = 3.5$ ML/s. In the absence of exchange
processes ($\Ux = \infty)$ both anisotropy parameters are smaller than
zero for all temperatures, leading to a preference of in-plane
magnetization. The reason is that the clusters are relatively flat and
that surface segregation of Pt is weak because of kinetic suppression.
As a consequence, in the surface of a cluster there exist more Co-Pt
bonds in-plane than out-of-plane, while atoms in the inner part of the
cluster give essentially no contribution to $P$. The flat cluster
shape, on the other hand, would favor PMA if strong Pt segregation as
realized at equilibrium could build up. 

Indeed, when including exchange processes, Pt segregation gets
enhanced considerably and $P$ can become positive. Actually, the
cluster displayed in Fig.~\ref{fig:cluster-shape}, showing pronounced
Pt segregation to the (111) and (100) facets, was computed with
$U_{\rm x} = 0$. As seen from Fig.~\ref{fig:struct_anisotropy-T}a, for
$U_{\rm x} = 0$, $P^{\rm CoPt}$ changes sign near $ T = 0.58$ and
reaches a maximum at an optimum temperature $T_{\rm max} \simeq 0.8$.
Conversely, $P^{\rm CoCo} < 0$ for all parameters, which leads to a
small reduction of the sum (\ref{Es}) relative to its leading term
$\alpha = $Pt, see Fig.~\ref{fig:struct_anisotropy-T}b.

The third set of data in Fig.~\ref{fig:struct_anisotropy-T}
(triangles) refers to $\Ux = 5$. This means that the activation energy
for direct exchange is twice the diffusion barrier. Yet exchange
processes have an important influence as they still can render $P^{\rm
  CoPt}$ positive. The onset of a positive $P^{\rm CoPt}$ and the
temperature were $P^{\rm CoPt}$ takes its maximum are shifted to
somewhat higher values than in the case $\Ux = 0$.
Fig.~\ref{fig:struct_anisotropy-T} also contains data for the flux $F
= 0.35$ ML/s. The reduction of $F$ by one order of magnitude
apparently leads to a small increase of $P^{\rm CoPt}$.

Investigating the mechanism of PMA in more detail we find that the
sign of $P^{\rm CoPt}$ is determined by two major factors which show
opposing trends in their temperature dependence.  These are the degree
of Pt surface segregation and the cluster shape, which we now discuss
in more detail.

\begin{figure}
\begin{center}
  \includegraphics[width=9cm]{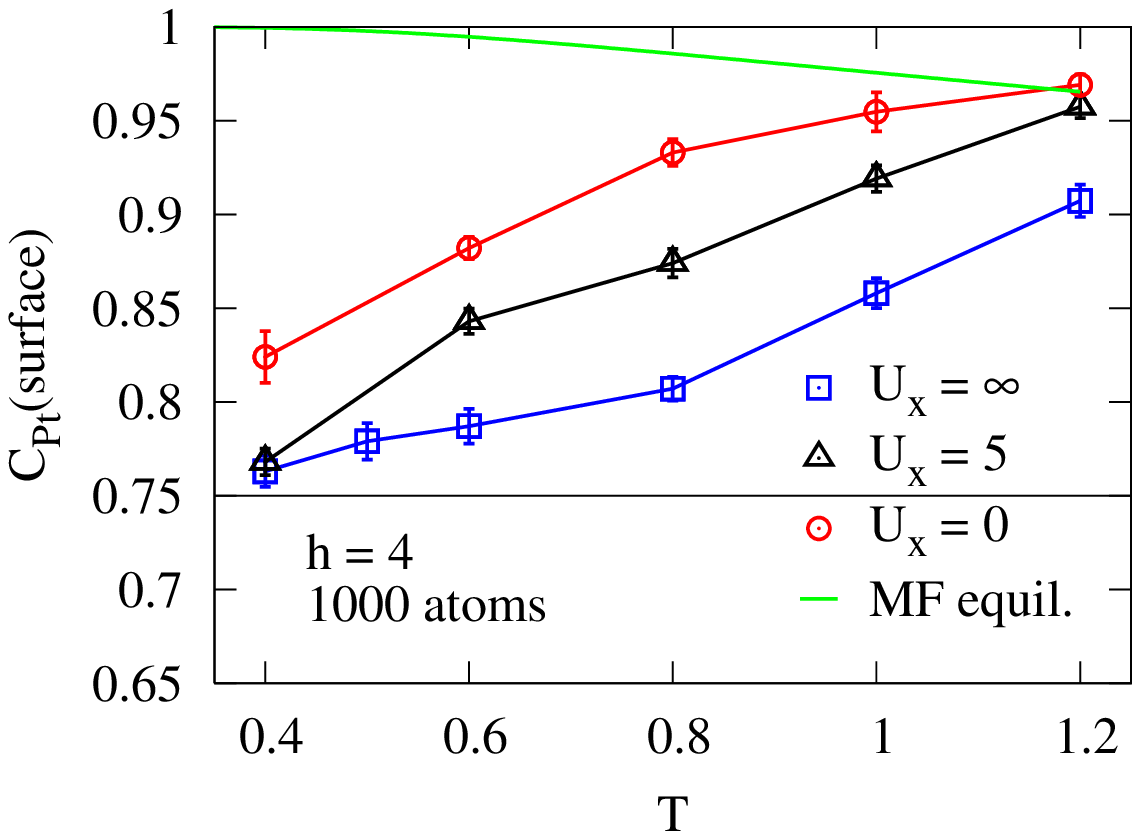}\\
  \includegraphics[width=9cm]{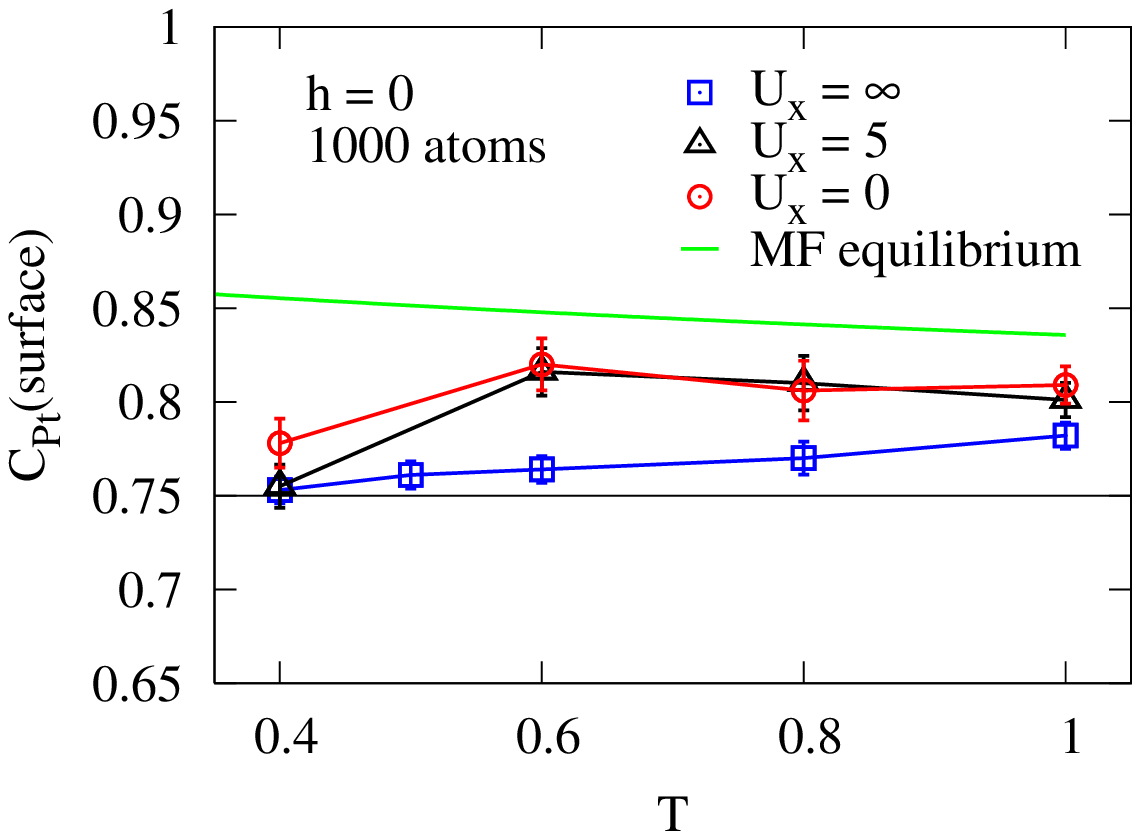}
\end{center}
\caption{Concentration of Pt atoms in
  the outer shell of clusters for $h=E_{\rm PtPt}-E_{\rm CoCo}=4$ in
  the upper panel and $h=0$ in the lower panel. The uppermost lines
  result from the mean field equations (\ref{eq:MF-U}) and
  (\ref{eq:MF-S}).}
\label{fig:segregation}
\end{figure}
The concentration $C_{\rm Pt}$ of Pt-atoms in the outer shell of a
1000 atom cluster is plotted in Fig.~\ref{fig:segregation}, again for
a flux $F = 3.5$ ML/s and three values of $\Ux$. As discussed above,
the observed degree of Pt surface segregation is generally smaller
than in the equilibrium case because of kinetic hindrance: During
growth, the time for attaining equilibrium through exchange and
diffusion processes is limited due to continuous incorporation of
newly deposited atoms. Fig.~\ref{fig:segregation} reveals that both
direct exchange processes and an increasing temperature act towards
restoring equilibrium, i.e.\ they facilitate Pt surface segregation.

Clearly, a large $C_{\rm Pt}$ in the topmost layer leads to an
enrichment of Co-atoms in the layer underneath. Therefore, for oblate
cluster shapes with more surface area oriented in the [111] direction
than in directions perpendicular to [111], a sufficiently strong
segregation of Pt to the surface will induce a positive sign of
$P^{\rm CoPt}$. This favors PMA, even for a disordered structure
within the cluster interior. Furthermore, the weakly binding van der
Waals substrate also allows for segregation towards the substrate
during growth, effectively doubling the available surface with
favorable orientation.

It should be noted that segregation of the majority atoms (Pt) can be
observed even when the parameter $h = V_{\rm PtPt} - V_{\rm CoCo}$
controlling segregation is zero. This can be understood from a
mean-field argument that counts bonds in a slab with a free (111)
surface: for a random fcc alloy structure, the exchange of a Co-atom
at the surface with a Pt-atom in the bulk allows the system to lower
its energy on average by 3$I$, so that the surface concentration
$C_{\rm Pt}$ will exceed the stoichiometric concentration, see
Fig~\ref{fig:segregation}. By contrast, for a fully L1$_2$ ordered
alloy the stoichiometric composition at the surface yields the lowest
energy.

In order to give a more quantitative estimate of segregation, we
consider a mean field model that consists of three completely filled
(111) layers of atoms. The first layer is the free surface and the
third one has fixed stoichiometric composition. Exchange processes are
allowed between the outermost two layers. This model takes into
account the almost non-existent bulk diffusion within a cluster, which
results in an increased concentration of Co in the second layer
impeding further exchange. The internal energy term per atom for a
given Co concentration in the first layer $C_{\rm Co,1}$ can be
expressed in the case of 1:3 stoichiometry as
\begin{equation}
U=6 V_0 - \frac{3}{16}h\,(1-4\,C_{\rm Co,1}) + \frac{3}{2} I\,(1-C_{\rm
  Co,1}+4\,C_{\rm Co,1}^2).
\label{eq:MF-U}
\end{equation}

The energy minimum at $T=0$ is attained for $C_{\rm
  Co,1}=\frac{1}{8}(1-h/(2J))$ clearly predicting segregation for
$h=0$.

The entropy is
\begin{equation}
S=-\frac{\kB}{2} \sum_{\rm \alpha=Co,Pt}\;\sum_{z=1}^2 C_{\alpha,z}\ln C_{\alpha,z}.
\label{eq:MF-S}
\end{equation}
Minimizing the free energy $F=U-TS$ then yields the equilibrium
concentration of Co and Pt in the first layer. The corresponding
results for the Pt concentration are displayed in
Fig.~\ref{fig:segregation}. Including the entropy term leads to a
decrease in Pt concentration in the outer shell with increasing
temperature and restores the stoichiometric concentrations in the
limit of high temperatures. Such a negative slope with temperature is
reproduced in the simulations for $h=0$ when exchange processes are
included, but not in the simulations for $h=4$ and $h=0$ without
exchange processes. The positive slope of simulated segregation values
with temperature indicates that segregation differs from thermal
equilibrium due to kinetic limitations. This model also does not
account for the portions on the cluster surface which are not flat.

\begin{figure}
\begin{center}
  \includegraphics[width=8cm]{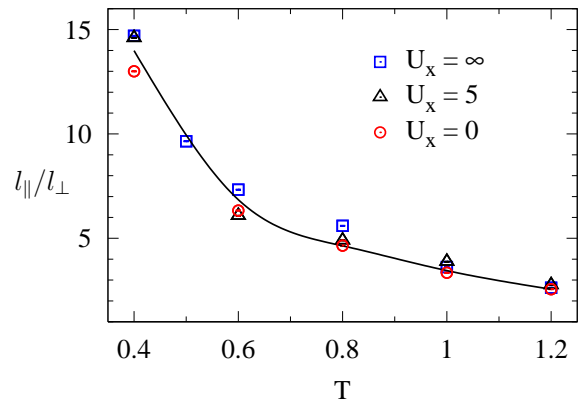}
\end{center}
\caption{Cluster shape represented by the ratio of gyration radii
  $l_\parallel/l_\perp$ (with line as guide to the eye). Decreasing
  values signify the transition from oblate to spherically shaped
  clusters.}
\label{fig:shape_parameter-T}
\end{figure}
\begin{figure}
\begin{center}

  \includegraphics[width=9cm]{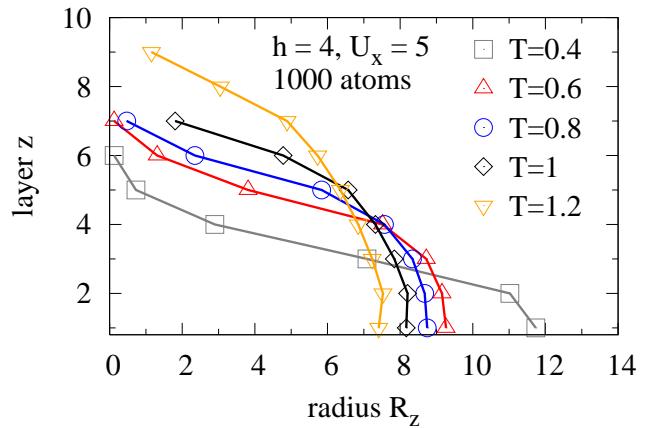}
\end{center}
\caption{Number of atoms per layer $N_z$ represented with a radius
  $R_z=\sqrt{N_z/\pi}$ and averaged over 20 realizations of the growth
  process. Layer one is the bottom layer on the (111) substrate.}
\label{fig:atoms-layers}
\end{figure}

The second factor important for PMA is the cluster shape. An
increasing temperature drives the shape closer to the equilibrium
shape which is less oblate. This is seen from the aspect ratio
$l_\parallel/l_\perp$ of the gyration radii of cluster sizes in the
direction parallel and perpendicular to the substrate, shown in
Fig.~\ref{fig:shape_parameter-T}. Cluster shapes are parameterized in
more detail in Fig.~\ref{fig:atoms-layers}. The atomic layer $z$
contains $N_z$ atoms and has an effective radius $R_z =
\sqrt{N_z/\pi}$. The figure displays the connection between $z$ and
$R_z$ for a series of temperatures. It clearly shows the evolution of
cluster shapes from oblate to almost a half-sphere as temperature is
increased. In this way temperature effects in the cluster shape
counteract the temperature dependent, segregation induced PMA. The
result is the existence of a certain temperature window where PMA
prevails, in agreement with experiments.\cite{Albrecht+02,Albrecht+01}

\begin{figure}
\begin{center}
  \includegraphics[width=8cm]{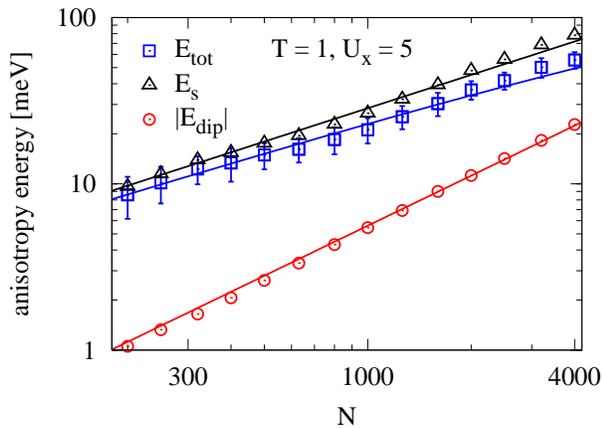}
\end{center}
\caption{Scaling plot of surface $E_{\rm s}$ and bulk $E_{\rm dip}$ contributions
  to the magnetic anisotropy $E_{\rm tot}=E_{\rm s}+E_{\rm dip}$ as a
  function of cluster size $N$. The lines fitting $E_{\rm s}$ and
  $E_{\rm dip}$ have slope 2/3 and 1 respectively. The line through
  $E_{\rm tot}$ is given by $E_{\rm tot}=K_{\rm s} N^{2/3}-K_{\rm dip}
  N$, where the prefactors $K_{\rm s}$ and $K_{\rm tot}$ are obtained
  from the fits to $E_{\rm s}$ and $E_{\rm dip}$.}
\label{fig:magneticE-scaling}
\end{figure}

\begin{figure}
\begin{center}
  \includegraphics[width=8.5cm]{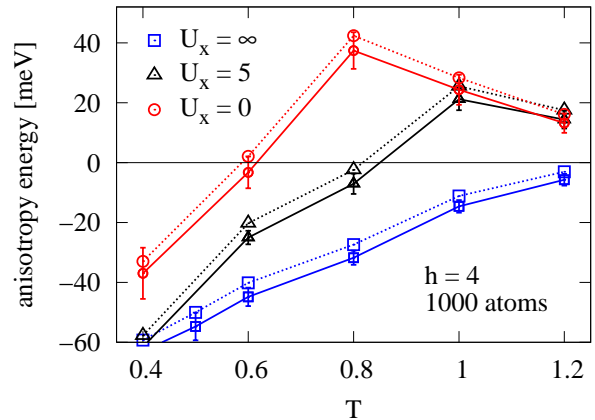}\end{center}
\caption{Magnetic anisotropy energies $E_{\rm tot}$ (solid lines) and
  $E_{\rm s}$ (dashed lines) as a function of temperature for
  different values of the additional barrier $\Ux$ for exchange
  processes (no exchange processes for $\Ux=\infty$).  $E_{\rm tot}$
  also includes the form anisotropy from dipolar interactions, while
  $E_{\rm s}$ only contains the bond contributions (with inclusion of
  the nearest neighbor part of the dipolar interactions).}
\label{fig:magnetic_anisotropy-T}
\end{figure}

The discussion so far makes it clear that PMA essentially emerges as a
surface effect. This suggests that $n_\perp^{\rm CoPt} - n_\parallel
^{\rm Co Pt} \sim N^{2/3}$, and accordingly $P^{\rm CoPt} \simeq
N^{-1/3}$. As shown by the solid line in
Fig.~\ref{fig:magneticE-scaling}, this behavior is well obeyed by the
simulated data (triangles).

Combination of our results for $P^{\rm Co \, \alpha }$ with (\ref{Es})
yields the structural part of the anisotropy energy. Adding the
dipolar energy (see section~\ref{sec:model}) we obtain the total
anisotropy energy (\ref{Etot}), which can be written as
\begin{equation}\label{Eaniso}
  E_{\rm tot} = K_s N^{2/3} - K_{\rm dip} N
\end{equation}
The temperature dependence of $E_{\rm tot}$ is shown in
Fig.~\ref{fig:magnetic_anisotropy-T} for different $\Ux$ and $N =
10^3$ atoms. An optimal temperature where $E_{\rm tot}$ is maximum can
clearly be identified. For example, for $\Ux = 5$ we have $T_{\max}
\simeq 1$. Fig.~\ref{fig:magneticE-scaling} contains a
double-logarithmic plot of $E_{\rm tot}$ as a function of $N$ at $T =
1, \, \Ux = 5$. From $E_{\rm s}$ and $E_{\rm dip}$ we obtain $K_{\rm
  s} \simeq 285\,\mu$eV and $K_{\rm dip} \simeq 5.6\,\mu$eV. The line
for $E_{\rm tot}$ in Fig.~\ref{fig:magneticE-scaling} represents the
expression (\ref{Eaniso}) and, upon extrapolation to larger
$N$-values, predicts the existence of an optimal cluster size for PMA,
which is $N_{\rm opt} \simeq 4\cdot 10^4$. PMA is expected to
disappear when $N \simeq 1.3 \cdot 10^5$.

Experimental values for $E_{\rm tot}$ for two different cluster sizes
can be inferred by measurements of the blocking temperature using
SQUID devices.\cite{Albrecht+01} For clusters with $N = 300$ at room
temperature $(T = 0.56)$, this leads to the estimate $E_{\rm tot}
\simeq 2.8\,$meV. Similarly, $E_{\rm tot} = 3.4$ meV for $N = 1200$ at
573 K ($ T = 1.1$). In a different set of experiments at $T = 0.56$,
granular nanostructures (dense covering of surface, but not touching)
with similar lateral size of 3 nm were obtained with considerably
larger anisotropy constants $K \simeq 13\,\mu$eV per
atom.\cite{Albrecht+01} In our simulation model values $E_{\rm tot}
\lesssim 12\,$meV for $N = 300$ and $E_{\rm tot} \lesssim 20\,$meV for
$N = 1200$ are found for $T \gtrsim 0.9$.  The comparison shows that
the experimentally observed values of the anisotropy energy lie within
the ranges predicted by the model. The degree of agreement with
experiment must be regarded as satisfactory in view of the
uncertainties in the analysis of experiments and the simplifying
assumptions in our model. One should also note the slower deposition
rate in the experiments with $F = 0.02$ ML/s, compared to $F = 3.5$
ML/s used in most simulations. An effect of slower deposition rates in
the simulations can be seen in Fig.~\ref{fig:struct_anisotropy-T}.

Another effect to be mentioned is that larger spin-orbit couplings are
known to occur for Co-atoms with low coordination on terraces or along
steps on Pt-surfaces.\cite{Gambardella+05} Such an effect in principle
can further enhance PMA in nanoclusters because of the flat cluster
shape. However, because of strong Pt surface segregation in the
nanoclusters most of the near-surface Co-atoms are buried in the
second layer or have high coordination. Hence we expect that the few
Co-atoms found in the outermost layer need not be treated separately.

\begin{figure}
\begin{center}
  \includegraphics[width=8cm]{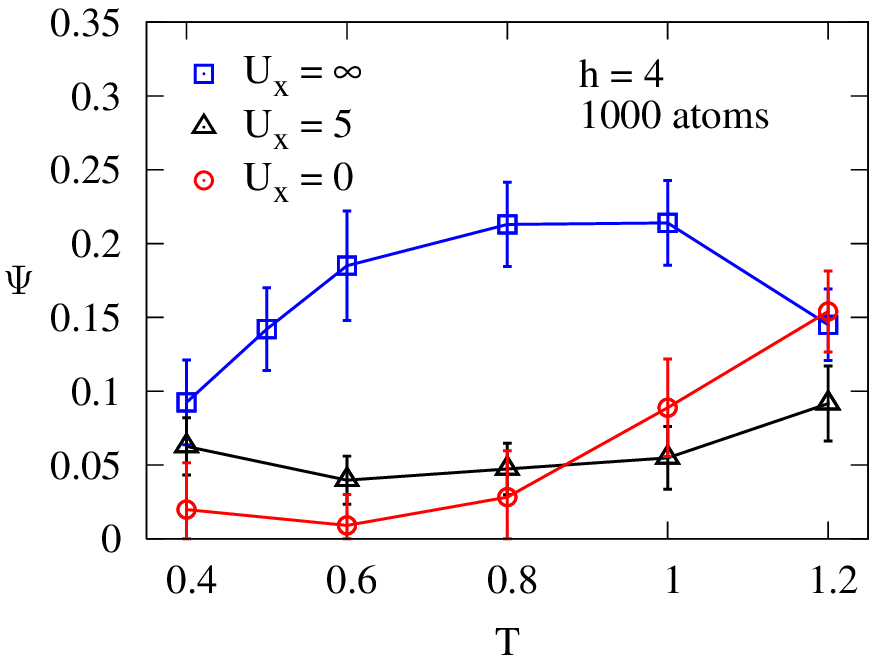}\\[1cm]
  \includegraphics[width=8cm]{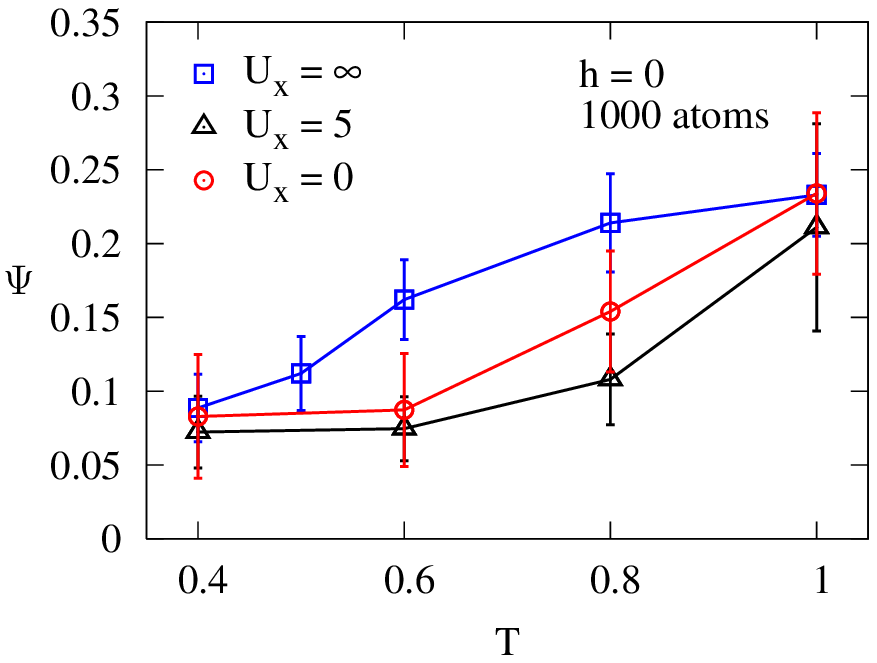}
\end{center}
\caption{Dependence of the order parameter for L1$_2$ structure $\Psi$
  on temperature for segregation controlling parameter $h=4$ (upper
  panel) and $h=0$ (lower panel). Values are averaged over 20
  realizations for $h=4$ and 10 realizations for $h=0$.}
\label{fig:orderpar-T}
\end{figure}

Next we turn to analyzing the L1$_2$-ordering of clusters. Analogous
to experiments, we determine the L1$_2$ order parameter from the
magnitudes of scattered intensities around the associated three
superstructure peaks $\Kv_i$, $i = 1,2,3$ in reciprocal space. For
that purpose, atoms of the cluster and the surrounding vacancies are
represented by pseudo-spins $s_l \in \{1,0,-1\}$ at the lattice
positions $\Rv_l$ with $s_l = 0$ for vacancies and $s_l = \pm 1$ for
$A$ and $B$, respectively. The structure factor is then calculated
from the amplitudes $F_\kv = \sum_l s_l e^{-i\kv\cdot \Rv_l}$ which account
for both the atomic arrangement in the cluster and the cluster shape.
The finite dimensions of the cluster lead to a significant broadening
of peaks in Fourier space, especially in the vertical direction. In
order to account for this broadening, an integration in $\kv$-space
was performed around each peak in form of a sphere with radius
0.1$/a$. The total scattering intensity is calculated as $ I =
\sum^3_{i=1} \sum _{|\kv - \Kv_i| < 0.1/a} |F_\kv|^2 $. As an
order parameter we define
\begin{equation}\label{orderpara}
  \Psi = \frac{I - I_{\rm random}}{I_{{\rm L1}_2}- I_{\rm random}}
\end{equation}
Here $I_{L1_2} $ and $I_{\rm random}$ are the intensities of clusters
with identical shape, but perfect L1$_2$ order and random occupation
by A and B atoms, respectively.

Fig.~\ref{fig:orderpar-T} shows the temperature dependent order
parameter $\Psi$ in cases of a strong surface field $h = 4$ (a) and $h
= 0$ (b). The inclusion of exchange processes ($\Ux < \infty$)
diminishes $\Psi$ due to reduced ordering through segregation effects.
This can be understood from the large energy a Co adatom gains when it
is incorporated by an exchange process into deeper layers of the
cluster irrespective of its contribution to ordering.

In the presence of exchange processes we notice from the figures that
ordering sets in at temperatures near $T \simeq 0.8$ to 1 (420 --
523$\,$K). In experiments, the onset of L1$_2$ ordering was found for
temperatures near 423$\,$K for 3$\,$nm thick films consisting of a
dense assembly of islands.\cite{Albrecht+02} This temperature range is
somewhat lower but close to the corresponding range in the
simulations.

Comparison with Fig.~\ref{fig:magnetic_anisotropy-T} shows that PMA
for $\Ux = 5$ starts to decrease for $T \gtrsim 1$, which coincides
with the onset of L1$_2$ ordering. One should bear in mind, however,
that PMA in clusters is a surface induced effect and its decrease is
caused by the change in cluster shape rather than by the concomitant
bulk ordering.

\begin{figure}[b]
\begin{center}

  \includegraphics[width=9cm]{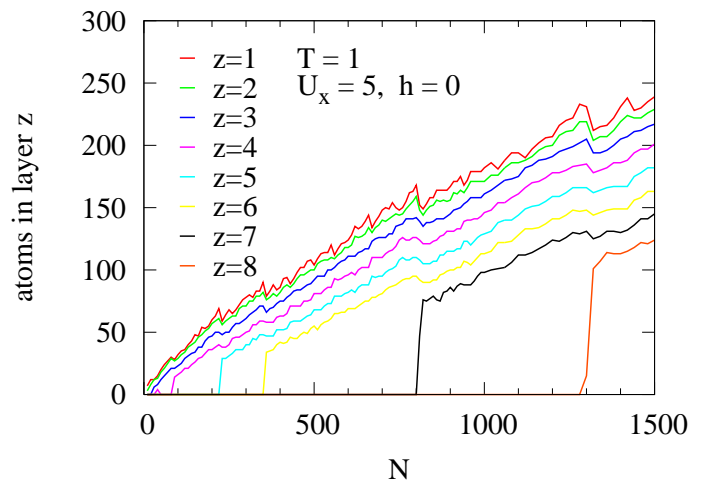}\end{center}
\caption{Atoms per layer versus total number of atoms during the growth of
  one cluster. The curve on top shows the number of atoms in the first
  layer above the substrate and curves below represent successive
  layers on top.  For the 4th to 8th layer nucleation events are
  visible as a strong initial increase in the number of atoms. }
\label{fig:cluster-layer-growth}
\end{figure}

To better visualize the statistics of growth, we present in
Fig.~\ref{fig:cluster-layer-growth} the layer-resolved evolution of
one cluster. One can see pronounced incubation periods, before a
nucleation event on top of the last layer takes place. Such an event
is accompanied by a rapid initial growth and completion of the newly
formed layer, followed by a period of essentially lateral growth of
the whole cluster without change in height. The slope in the atom
numbers of the top layer immediately after nucleation is much higher
than one. This means that most of the atoms incorporated in a new
layer during its completion arrive through mass transport along the
side facets, leading to a transient lateral shrinkage of all the
layers below the growing top layer. This is seen in the figure by the
small dips in all curves occurring simultaneously during the short
time intervals when the top layer is filled.

The nucleation events on top can be described by a theory originally
devised for second layer nucleation.\cite{Heinrichs+02} This theory
describes nucleation in a confined geometry with an influx of
particles through deposition and loss of particles over a step edge.
An essential parameter is the Schwoebel barrier that atoms have to
surmount when crossing the step edge. In our simulations, this extra
energy barrier effectively is about $\DEs=V_0$ because the
intermediate state for a transition from the top facet to a side facet
has one bond less than adatoms with coordination three on the top
surface. With the parameters typically used in the simulations, the
fluctuation dominated regime III discussed in
Ref.~\onlinecite{Heinrichs+02} is the relevant one for critical nuclei
of size $i=1,2$. However, with the extra influx of atoms from the side
facets, it is likely that the mean number of adatoms on top of the
facet satisfies $\bar n>i$, so that mean-field theory may become
justified for top layer nucleation.\cite{Heinrichs+02}

\section{Films}
\label{sec:films}
\begin{figure}
\begin{center}
  \includegraphics[width=8cm]{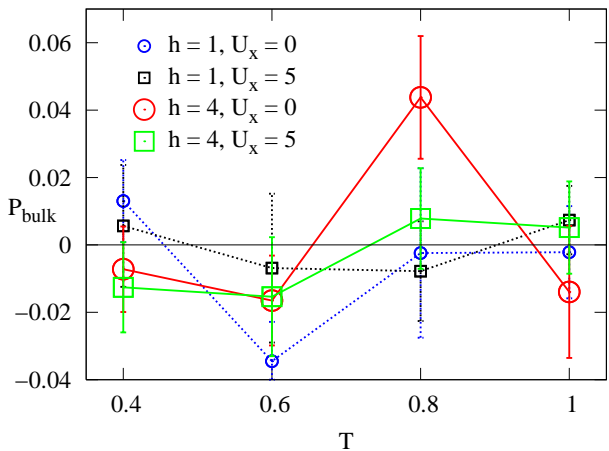}
\end{center}
\caption{Anisotropy parameter $P_{\rm bulk}$ for films as a function of
  temperature for different values of the exchange energy barrier
  $U_{\rm x}$ and surface field $h$. The error bars are calculated
  from the statistical error after averaging over 5 realizations of
  the growth process and the fit uncertainty ($P$ as a function of
  $1/N$, see text) with error propagation.}
\label{fig:film-anisotropy}
\end{figure}

The same model is now applied to continuous films. We simulate the
growth of films starting with an empty simulation box with 224 atoms
per layer and simulate growth up to deposition of 29 layers. The
surface binding energy of $V_s=3 V_0=-15$ induces film instead of
cluster growth. As before, the surface binding is independent of the
atom type so that segregation to the substrate similar to a free
surface can be observed. Consequently, true bulk properties are only
found a few layers above the substrate and below the top surface. In
order to separate the bulk contribution from surface effects, we
considered each realization of the growth process in the range of
2000 to 6500 atoms and made a linear fit of $P$ vs. $1/N$. The
anisotropy parameter in the bulk $P_{\rm bulk}$ is then found by
extrapolating $N\to \infty$.\cite{comm-P_bulk}
Fig.~\ref{fig:film-anisotropy} shows the temperature dependence of the
anisotropy parameter $P_{\rm bulk}$ for different values of $\Ux$ and
$h$.  For most sets of parameters, $P_{\rm bulk}<0$, i.e.\ Co-Pt bonds
align preferentially in the film plane. However, for certain
combinations of parameters, the simulations yield $P_{\rm bulk}>0$,
supporting PMA.

Clearly, even if $P_{\rm bulk}>0$ it is too small to account for the
measured PMA in thick CoPt$_3$ films at elevated temperatures. As
shown recently, those measurements can be explained by a model with
interatomic interaction potentials that depend on the coordination of
the atoms involved.\cite{Maranville+06} One essential feature of this
model is that an effective segregation parameter, related to $h$ in
our model, changes sign for intermediate coordinations of Co-atoms and
thus favors Co segregation towards step edges in the surface. This
sign change of $h$, however, should not significantly affect our
conclusions. There we focused on the temperature range $T \lesssim 1$,
where Co-segregation according to Ref.~\onlinecite{Maranville+06}
ceases to be effective.

\begin{figure}[t]
\begin{center}
  \includegraphics[width=8cm]{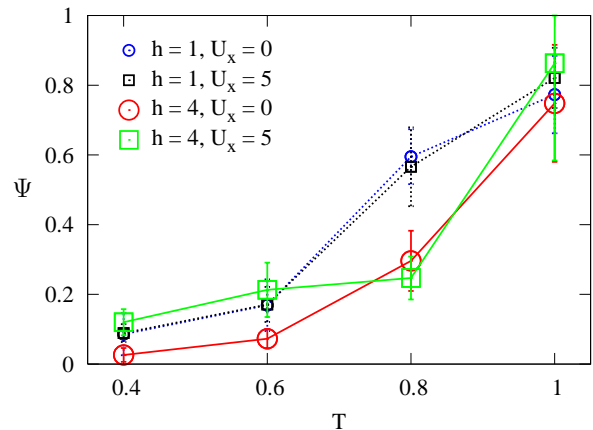}
\end{center}
\caption{Long range order parameter $\Psi$ for films as a function of
  temperature for different values of the exchange energy barrier
  $U_{\rm x}$ and surface field $h$. Values are averaged over 5
  realizations of the growth process.}
\label{fig:film-order}
\end{figure}

The degree of long-range order in films is shown in
Fig.~\ref{fig:film-order} at temperatures below the transition
temperature $T_0 \simeq 1.83$, where bulk kinetics are suppressed.
With higher $T$, the order parameter $\Psi$ becomes larger. Similar to
our findings for clusters, the order parameter $\Psi$ in the case
$h=4$ is suppressed relative to $h=1$, as clearly seen from the data
at $T=0.8$. Surface segregation therefore appears to impede the
development of long range order.

\section{Conclusions}
\label{sec:conclusions}

In order to study growth kinetics of binary alloys we developed a
lattice model based on nearest neighbor bonds with bond energies
chosen to match equilibrium properties of CoPt$_3$. The kinetic
parameters were obtained from diffusion experiments,\cite{Bott+96} DFT
calculations,\cite{Feibelman99_PRB} and in addition from experiments
observing interatomic exchange processes in the Co/Pt
system.\cite{Gambardella+00,Lundgren+99}

Experiments on CoPt$_3$ nanoclusters on a weakly binding
substrate\cite{Albrecht+01} have revealed PMA in a temperature window
that is well reproduced by our simulations. This temperature range is
bounded towards low temperatures by frozen-in surface kinetics. The
disappearance of PMA at higher temperatures is explained by the
interplay of Pt surface segregation facilitated by direct exchange
processes, and a transition from oblate to spherical cluster shapes.
Our analysis suggests that the transition is not caused by L1$_2$
ordering. Yet, in our simulations the onset of L1$_2$ ordering is
detected in the same temperature range where PMA disappears, in
qualitative agreement with the measurements. It should be noted that
up to $T \simeq 1$ long range order is induced solely by surface
processes, as the bulk kinetics remain still frozen.

The structural anisotropy responsible for the magnetic anisotropy is
characterized in our model primarily by a difference in the numbers of
Co-Pt bonds out-of-plane and in-plane. Within a bond picture each
Co-Pt bond provides a local contribution to the magnetic anisotropy,
which tends to align the Co-moment parallel to the bond. An associated
magnetic bond energy, deduced from experiments on Co-Pt multilayers,
can reasonably reproduce the magnitude of PMA measured for
nanoclusters in the appropriate temperature range. PMA is predicted to
be a surface effect, a feature which could be tested experimentally.

Dipolar interactions tend to turn the easy axis of magnetization into
the plane. The two competing effects, bond anisotropy (surface term)
and dipolar interactions (volume term), lead to an optimum cluster
size where the anisotropy energy of a cluster is largest, and a second
characteristic size where PMA disappears.

The importance of cluster shape effects revealed by our computations
is also corroborated by experiment, where a rotation of the easy
magnetization axis into the film plane and an associated change in the
aspect ratio from oblate to prolate was found in separate measurements
on cluster assemblies at higher coverages.\cite{Albrecht+02} In our
simulations, we observed clusters to become more spherical in shape
and to develop an excess of in-plane Co-Pt bonds with increasing
temperature.

Application of the same methodology to growing continuous films yields
a bulk structural anisotropy favoring in-plane magnetization for most
of the parameters studied. However, it can yield a positive bulk
contribution to $P^{\rm CoPt}$ in a certain temperature range provided
direct exchange processes are sufficiently fast. The associated PMA,
however, is quite small and disappears for $T \gtrsim 1$. At these
higher temperatures a different mechanism has been proposed recently
to describe PMA in thick films in terms of in-plane Co
clustering.\cite{Maranville+06}

Finally, we like to draw attention to the question of growth of
magnetic clusters or films in strong external magnetic fields that
saturates the magnetization in the growth direction. The magnetic
anisotropy energy should then induce an anisotropy in the
probabilities for atomic hopping such that out-of-plane Co-Pt bonding
and PMA become favored. In fact, simulation results for growth in a
magnetic field suggest that this effect might become detectable in the
Co-Pt system. In particular, for L1$_0$ ordered alloys estimates based
on Landau theory suggest that this additional field-induced PMA
becomes stronger.\cite{Einax+06}

\begin{acknowledgements} 
  We thank M. Albrecht, M. Einax, M.  Kessler, A. Maj\-hofer and G.
  Schatz for helpful discussions. This work was supported in part by
  the Deutsche Forschungsgemeinschaft (SFB 513).
\end{acknowledgements}



\begin{thebibliography}{44}
\bibitem{comm-A_par} The value for $A^{\rm CoPt}$ can be afflicted
  with uncertainty arising from measurements of films with varying
  interface quality through different preparation methods. In a
  previous work\cite{Heinrichs+06} we estimated this parameter based
  on experiments with values $K^{\rm CoPt}$ falling in the
  range\cite{Harzer+92,McGee+93} of 0.27--1.15~mJ/m$^2$.  The model
  can also describe the major contribution of the anisotropy resulting
  from alternating Pt and Co atomic layers in the ordered L1$_0$-phase
  of CoPt, where the lattice distortion (tetragonalization) is only
  responsible for about 20\% of the observed
  anisotropy.\cite{Razee+01_PRB}
\bibitem{comm-P_bulk} An alternative determination of $P_{\rm bulk}$
  where the outermost two layers at the top and bottom are excluded
  yields comparable results. We defined the outermost two layers to
  consist of atoms with coordination less or equal to 11 together with
  their nearest neighbors.
\end{thebibliography}
\end{document}